\documentclass{ifacconf}
\usepackage{amsfonts,amsmath,amssymb}
\usepackage{graphicx}      
\usepackage{natbib}        
\usepackage{subcaption}
\usepackage{verbatim}
\usepackage{mathbbol}
\usepackage{subcaption}
\newtheorem{lemma}{\textbf{Lemma}}

\newtheorem{corollary}{\textbf{Corollary}}
\newtheorem{definition}{\textbf{Definition}}
\newtheorem{remark}{\textbf{Remark}}
\newtheorem{expm}{\textbf{Example}}
\newtheorem{assump}{\textbf{Assumption}}
\begin{document}

\begin{frontmatter}

\title{Topology-based Conditions for Multiconsensus under the Signed Friedkin-Johnsen Model}

\thanks[footnoteinfo]{Sponsor and financial support acknowledgment
goes here. Paper titles should be written in uppercase and lowercase
letters, not all uppercase.}

\author[1]{Aashi Shrinate, Tanmay Siddharth, Twinkle Tripathy} 

\address[1]{Indian Institute of Technology, Kanpur, India,(e-mail: aashis21@iitk.ac.in,tanmays22@iitk.ac.in ttripathy@iitk.ac.in)}

\begin{abstract}    
In this paper, we address the multiconsensus problem in networked systems, where agents are partitioned into disjoint subgroups and the states of agents within a subgroup are driven to consensus. Our objective is to present a distributed control law that leads to multiconsensus in signed digraphs. To this end, we examine the convergence of opinions under the opposing rule-based signed Friedkin-Johnsen (SFJ) model and present conditions that lead to multiconsensus under this model. Interestingly, the proposed conditions depend only on graph topology and signed interactions and not on the edge weights of the network. Consequently, the proposed SFJ-based control law relaxes the \textit{in-degree balance} and \textit{homogeneity of trust-distrust}, frequently assumed in the literature. Finally, we add simulation results to demonstrate the proposed conditions for multiconsensus.





\end{abstract} 


\end{frontmatter}

\section{Introduction}

Emergent behaviours in networked systems are shaped by the interplay between the network topology and agent dynamics. 
Consensus among agents in a network has been widely explored due to its wide-ranging applications in engineered \cite{} as well as social networks \cite{LWang_CFJ}. Multiconsensus (group consensus) is another special emergent behaviour which occurs the agents in a network segregate into disjoint subgroups (clusters), such that the agents within each subgroup converge to the same steady-state value. In engineered networks, multiconsensus has numerous applications, such as task allocation in robot swarms \cite{Switching_T}, convergence to multiple rendezvous points \cite{tomaselli2023multiconsensus}, and in decision making.  In social networks, multiconsensus represents the existence of diverse views with subgroups of individuals with aligned views on a topic.

\textit{Related Literature:} Recent works, such as \cite{Yu_wang,xia2011clustering,qin2013group,luo2021cluster,maria}, propose several control techniques to achieve multiconsensus in networked systems. In \cite{Yu_wang,xia2011clustering}, the authors use pinning control to achieve multiconsensus in a network of agents with single-integrator dynamics. To prevent the opinions of agents in two different clusters from becoming equal, negative interactions between agents in different clusters are introduced in \cite{xia2011clustering}, while agents within a cluster interact cooperatively. Additionally, these works consider that the magnitude of cooperative intracluster coupling is sufficiently high to counter the effects of negative intercluster interactions. This requirement is relaxed in \cite{QIN20132898}, where it is shown that if the graph can be partitioned such that the clusters form an acyclic graph, then the pinning control technique leads to multiconsensus. This study further generalised to include such partitions of the network that do not form acyclic graph in \cite{luo2021cluster}. However, the works \cite{Yu_wang,xia2011clustering,QIN20132898} and \cite{luo2021cluster} consider that the `in-degree balance condition' holds, meaning that at each node the weights of incoming edges from another cluster of cooperative and antagonistic nature balance  each other.

 \textit{Structurally balanced} graphs, introduced in \cite{cartwright1956structural}, is a popular class of signed networks. Under stuctural balance, the agents in a network get partitioned into two clusters such that intracluster interactions are cooperative, while intercluster interactions are always antagonistic. Such networks structures commonly occur in international relations \cite{askarisichani20201995},  social networks \cite{ilany2013structural} \textit{etc}. A similar notion of clustering balance was proposed in \cite{cisneros2020signed} for a signed network that are partitioned into more than two clusters. Since all intercluster interactions are antagonistic in networks with structural or clustering balance, they do not satisfy the in-degree balance condition. The authors in \cite{maria}  achieve multiconsensus in undirected signed networks with clustering balance when the condition of homogeneity of trust and distrust holds. Under this condition, each agent in a cluster distributes has a predefined positive weight among the cooperative intracluster interactions and a negative weight(s) among antagonistic intercluster interactions.  In this work, our objective is to extend the results presented in \cite{maria} by achieving multiconsensus in signed directed graphs while relaxing both the above-discussed conditions on the edge weights \textit{i.e.} the in-degree balance and homogeneity of trust-mistrust. 

In opinion dynamics, the Friedkin-Johnsen (FJ) model, proposed in \cite{friedkin1990opinions}, explains the occurrence of persistent disagreement in social networks. 
Under persistent disagreement, the opinions of different agents generally converge to distinct states. However, unlike multiconsensus, the opinions generally do not converge form opinion clusters at steady state. Rather, they converge arbitrarily, depending on the network topology and stubborn behaviour \cite{friedkin2015control}.  
More recently, the works \cite{yao2022cluster,shrinate2025opinionclustering} present topological conditions under which opinions of agents evolving under the FJ model achieve multiconsensus. These works demonstrate that the FJ model can serve as a distributed control law in multi-agent systems, leading to multiconsensus under suitable topology-based conditions.

The FJ model for signed networks was introduced in \cite{shrinate2025signed} based on the opposing rule (Refer to \cite{shi2019dynamics} for a description of the opposing and repelling rule). 
In this work, we show that the signed FJ model achieves multiconsensus in signed digraphs that do not satisfy `in-degree balance', thereby generalising the results in \cite{maria} to directed graphs. 
Interestingly, the proposed conditions are purely based on graph topology and sign of interactions but not on the weights of these interactions. Consequently, the homogeneity of weights required in \cite{maria} is also relaxed under the signed FJ framework.  
Our work is closely related to \cite{MONACO201953} and \cite{tomaselli2023multiconsensus} that use topological properties such as External Equitable Partitions and network symmetries, respectively, to define partitions in the network and achieve multiconsensus. However, unlike these works, we develop topological conditions that are also applicable to signed networks. Additionally, this work generalises the topological conditions developed in \cite{yao2022cluster,shrinate2025opinionclustering} for multiconsensus under the FJ model to signed networks.

\textbf{Paper Organisation:} Sec. \ref{Sec:Notations_Pre} presents the relevant notations and Preliminaries. Sec. \ref{Sec:Multiconsensus_in_SFJ} defines the multiconsensus problem for signed digraphs. The properties of Locally Topological Persuasive agents are presented in Sec. \ref{sec:LTP}. The conditions for multiconsensus  are presented in Sec. \ref{Sec:Main_Res} with simulation results. Finally, we conclude in Sec. \ref{Sec:Conclusion} with insights into future directions.

\section{Notations \& Preliminaries}
\label{Sec:Notations_Pre}
\subsection{Notations}
Let $\mathbb{N}$ and $\mathbb{R}$ denote the set of natural and real numbers, respectively. $\mathbf{1}$ denotes a vector or matrix of appropriate dimension and $I$ denotes the identity matrix of appropriate dimension. For any $n\in \mathbb{N}$, we denote the set of natural numbers $\{1,...,n\}$  as $[n]$. For a finite set $X$, its cardinality is given by $|X|$.
Consider a matrix $M=[m_{ij}] \in \mathbb{R}^{n \times n}$, then $\operatorname{abs}({M})$ is the matrix with the $(i,j)^{th}$ entry equal to $|m_{ij}|$ $\forall \ i,j \in \{1,...,n\}$. The spectral radius of $M$ is denoted by $\rho(M)$.  
\subsection{Graph Preliminaries}
A directed graph (digraph) is denoted as $\mathcal{G}=(\mathcal{V},\mathcal{E})$ with $\mathcal{V}$ being the set of $n$ agents and $\mathcal{E} \subseteq \mathcal{V} \times \mathcal{V}$ giving the set of directed edges in the digraph. An edge $(i,j) \in \mathcal{E}$ implies that node $i$ is an in-neighbour of node $j$ and node $j$ is an out-neighbour of node $i$. The matrix $A \in \mathbb{R}^{n \times n}$ is the adjacency matrix of a signed network, as the entries can take both positive and negative values. $A=[a_{ij}]$ is defined as: $a_{ij} \neq 0$ if $(i,j) \in \mathcal{E}$; otherwise $a_{ij}=0$.

A walk in a graph $\mathcal{G}$ is a sequence of nodes where each consecutive pair of nodes forms an edge. A path is a walk that does not contain repeated nodes. The weight of a walk (path) is the product of the edge weights of all the edges along the walk (path).

\subsection{Matrix Preliminaries}
Consider a matrix $M\in \mathbb{R}^{n \times n}$ such that $\rho(M)<1$. By Neumann series, $(I-M)^{-1}=\sum_{k=0}^{\infty}M^k$.

For a matrix $M\in \mathbb{R}^{n \times n}$, its associated digraph $G(M)$ has $n$ nodes and an edge $(j,i)$ exists if the entry $m_{ij}$ is non-zero for all $i,j\in[n]$.

Consider a matrix $M\in \mathbb{R}^{n \times n}$. Let $\alpha,\beta\subseteq[n]$ be index sets. Then, the submatrix of $M$ with rows indexed by $\alpha$ and columns indexed by $\beta$ is denoted by $M[\alpha,\beta]$. Additionally, submatrix $M[\alpha,\alpha]$  is simply denoted as $M[\alpha]$.
If $\alpha$ is an index set, then $\alpha^c=[n] \setminus \alpha$.  

Consider $\alpha^c$ such that $M[\alpha^c]$ is invertible. Then, the Schur complement of $M$ with respect to $M[\alpha^c]$ is given by:
\begin{align}
\label{eqn:Schur_complement}
    M/\alpha^c=M[\alpha]-M[\alpha,\alpha^c](M[\alpha^c])^{-1}M[\alpha^c,\alpha]
\end{align}

Schur complement is often used in solving linear equations by eliminating a block of variables. Consider a set of linear equations of the following form:
\begin{align}
\label{eqn:set_of_LE}
    \begin{bmatrix}
        M[\alpha] & M[\alpha,\alpha^c] \\
         M[\alpha^c,\alpha] &  M[\alpha^c]
    \end{bmatrix}\begin{bmatrix}
        \mathbf{x} \\ \mathbf{y}
    \end{bmatrix}=\begin{bmatrix}
        \mathbf{u} \\ \mathbf{v}
    \end{bmatrix}
\end{align}
where $\mathbf{x},\mathbf{u}\in \mathbb{R}^{|\alpha
|}$ and $\mathbf{y},\mathbf{v}\in \mathbb{R}^{n-|\alpha
|}$. By using the Schur complement, the eqns. \eqref{eqn:set_of_LE} reduce to 
\begin{align}
\label{eqn:reduced_LE}
   M/\alpha^c \mathbf{y}=\mathbf{v}-M[\alpha ,\alpha ^c]M[\alpha ^c]^{-1}\mathbf{u} 
\end{align}

Refer to \cite{zhang2006schur} for a detailed treatment of Schur Complement.

\begin{lemma}[\cite{shrinate2025signed}]
\label{lm:spectral_positive}
The spectral radius of a matrix $M \in \mathbb{R}^{n \times n}$ and $\tilde{M}=\operatorname{abs}(M)$, derived from $M$, satisfy $ \rho(M)\leq \rho(\Tilde{M})$.
\end{lemma}
\section{Multiconsensus in Signed Friedkin Johnsen opinion model}
\label{Sec:Multiconsensus_in_SFJ}
Consider a group of $n$ agents such that the interactions among these agents are both friendly (cooperative) and hostile (antagonistic) in nature. Antagonistic interactions occur naturally in social networks \cite{askarisichani20201995}, biological networks \cite{ilany2013structural} and are introduced in engineered networks to obtain desired collective behaviours \cite{Switching_T,QIN20132898}. A signed network $\mathcal{G}=(\mathcal{V},\mathcal{E})$ with $A\in \mathbb{R}^{n\times n}$ captures both cooperative and antagonistic interactions.  


In opinion dynamics, the FJ model is extensively used to model the opinions of agents in networks with cooperative interactions. A generalisation of the FJ model to signed networks was recently presented 
in \cite{shrinate2025signed} under the opposing rule. The proposed signed FJ (SFJ) model is given as follows: 
\begin{align}
\label{eqn:Signed_FJ_m}
    \mathbf{x}(k+1)=(I-\beta)W\mathbf{x}(k)+\beta \mathbf{x}(0)
\end{align}
here, $W=[w_{ij}]$ matrix is defined as:\begin{align}
\label{eq:weighted_Adjacency}
w_{ij}=\begin{cases}
    \frac{a_{ij}}{\sum_{j=1}^{n}|a_{ij}|} & \text{ if }\sum_{j=1}^{n}|a_{ij}| \neq 0  \\
   1 & \text{ if }\sum_{j=1}^{n}|a_{ij}| = 0 \text{ and } i=j\\
   0 & \text{ if }\sum_{j=1}^{n}|a_{ij}| = 0 \text{ and } i\neq j
\end{cases}    
\end{align}
By definition, each row of $W$ satsifies $\sum_{j=1}^{n}|w_{ij}|=1$ for $i\in [n]$. The matrix $\beta=\operatorname{diag}(\beta_1,...,\beta_n)$ is a diagonal matrix with entry $\beta_i\in[0,1]$ denoting the stubbornness of agent $i$. An agent $i \in \mathcal{V}$ is stubborn if $\beta_i>0$, otherwise it is non-stubborn. The emergent behaviours that arise under the SFJ model \eqref{eqn:Signed_FJ_m} due to the interplay between signed interactions and stubbornness were characterised in \cite{shrinate2025signed}. 


In this work, our objective is to present 
the conditions under which opinions of agents evolving by the SFJ model \eqref{eqn:Signed_FJ_m} achieve multiconsensus. Through this analysis, we establish that the SFJ opinion model can be used as a suitable distributed control law that leads to multiconsensus in signed networks. 



\subsection{Terminologies relating to Multiconsensus}
To begin our analysis, we present the following definitions.
\begin{definition}
 A digraph $\mathcal{G}=(\mathcal{V},\mathcal{E})$ with $n$ nodes admits a \textit{partition} if its nodes $\mathcal{V}$ can be split into $z \leq n$ non-empty subgroups $\mathcal{V}_1,...,\mathcal{V}_z$ that satisfy the following:
 \begin{itemize}
     \item $\cup_{i=1}^{z} \mathcal{V}_i=\mathcal{V}$,
     \item $\mathcal{V}_i \cap \mathcal{V}_j=\phi$ for all $i,j\in[n]$.
 \end{itemize}
\end{definition}

\begin{definition}
Consider a digraph $\mathcal{G}=(\mathcal{V},\mathcal{E})$ that admits a partition of agents in $z$ clusters (\textit{i.e.} $\mathcal{V}_1,...,\mathcal{V}_z$). If the final opinions 
of agents $i,j$ in each cluster $\mathcal{V}_h\subset \mathcal{V}$ satisfy:
\begin{align*}
   \lim_{k \to \infty} x_i(k)-x_j(k)=0 \quad \forall \ i,j\in \mathcal{V}_h, \quad \forall \ h\in[z]
\end{align*}
for any $\mathbf{x}(0) \in \mathbb{R}^n$.
Then, the agents in $\mathcal{G}$ achieve \textit{multiconsensus}. The agents in a partition that converge to the same final opinion are said to form an \textit{opinion cluster}.
\end{definition}

Note that under multiconsensus, the opinions of agents belonging to two distinct clusters may converge to the same opinion. Next, we examine the conditions under which multiconsensus can occur under the SFJ model \eqref{eqn:Signed_FJ_m}.

\subsection{Multiconsensus under SFJ model}
In this work, we consider that the network $\mathcal{G}$ satisfies the following assumption:
\begin{assump}
\label{Assumption:1}
Each non-stubborn agent in $\mathcal{G}$ has a path from a stubborn agent.
\end{assump}
Under Assumption \ref{Assumption:1}, it is demonstrated in (Theorem 1 in \cite{shrinate2025signed}) that $\rho((I-\beta)W)<1$ and the final opinions of agents converge as follows:
\begin{align*}
    \mathbf{x}^*=V \mathbf{x}(0)
\end{align*} 
where the matrix $V=[v_{ij}]$ is defined as: $V=(I-(I-\beta)W)^{-1}\beta$. 

A set of agents $\mathcal{C}\subseteq \mathcal{V}$ forms an opinion cluster under the SFJ model \eqref{eqn:Signed_FJ_m} if the entries of $V$ satisfy:
\begin{align}
\label{eqn:V_enteries}
v_{ih}=v_{jh} \qquad \forall i,j\in \mathcal{C},\forall \  h\in[n]
\end{align}

Since $\rho((I-\beta)W)<1$ under Assumption \ref{Assumption:1}, by the Neumann series, we can write $V$ as:
\begin{equation}
V=\sum_{k=0}^{\infty} ((I-\beta)W)^k\beta    
\end{equation}

Consequently, each entry $v_{ij}$ of $V$ is the product of $\beta_j$ with the sum of weights of all possible walks from $j$ to $i$ in the associated digraph $G((I-\beta) W)$. 
Therefore, multiconsensus occurs under the SFJ model \eqref{eqn:Signed_FJ_m}, if the network $\mathcal{G}$ has suitable edge weights such that the entries of $V$ satisfy the relation \eqref{eqn:V_enteries}. It is very tedious to design a network that satisfies eqn. \eqref{eqn:V_enteries}. Hence, the following question naturally arises:

\begin{enumerate}
    \item[Q-1] Can the topological properties of a signed network $\mathcal{G}$ ensure that eqn. \eqref{eqn:V_enteries} holds for a set of agents $\mathcal{C}\subset \mathcal{V}$?
\end{enumerate}

In this paper, by answering Q-1, we present the topological conditions that lead to multiconsensus under the SFJ framework.





\begin{remark}
Note that an SFJ model with the repelling rule was proposed in \cite{Rajaq}. Under this model, the opinions are shown to converge for $W$ that is eventually stochastic, meaning that $W\mathbb{1}_n=\mathbf{1}_n$ and there exists a $k_0$ such that $W^k>0$ for each $k\geq k_0$. Thus, the cooperative interactions of each agent must outweigh its antagonistic interactions and the underlying network must be strongly connected. Since we examine the topological properties of the networks for multiconsensus, we analyse the opposing rule-based SFJ model that converges for a larger class of signed digraphs (including the weakly connected digraphs).

\end{remark}
\section{Locally Topologically Persuasive (LTP) agents in Signed Networks}
\label{sec:LTP}
In \cite{shrinate2025opinionclustering}, we introduced the agents with special topological properties known as the LTP agents.
In a cooperative network, an LTP agent ensures the formation of an opinion cluster under the FJ model. In this work, we present the conditions under which LTP agents form opinion clusters in signed networks. We begin by stating the definition of LTP agents. 

\begin{definition}[\cite{shrinate2025opinionclustering}]
\label{Definition:LTP_agents}
Consider a network $\mathcal{G}=(\mathcal{V}, \mathcal{E})$ with $m$ stubborn agents labeled as $1,...,m$. Let an agent $p\in \mathcal{V}$ satisfy the following conditions:
\begin{enumerate}
    \item[(i)] a non-stubborn agent $q$ exists such that each path from every stubborn agent $s\in[m]$ to $q$ traverses $p$,
    \item[(ii)] if $p$ is non-stubborn, then there must not exist an agent $c \in \mathcal{V}$ such that the each path from stubborn agents to $p$ traverses $c$.
\end{enumerate}  
For an LTP agent $p$, a non-stubborn agent satisfying condition (i) is said to be persuaded by it. The set of all non-stubborn agents persuaded $p$ is denoted by $\mathcal{N}_p$. 
\end{definition}

\begin{remark}
An LTP agent can be either a stubborn or a non-stubborn agent. If an agent $p$ is stubborn, then it is sufficient to show that condition (i) in Defn. \ref{Definition:LTP_agents} holds. However, a non-stubborn LTP agent $p$ must additionally satisfy condition (ii) to ensure that another LTP agent $c$ does not exist such that $p\in \mathcal{N}_c$. 
\end{remark}

\begin{figure}[h]
    \centering
    \begin{subfigure}{0.22\textwidth}
        \centering
            \includegraphics[width=0.6\linewidth]{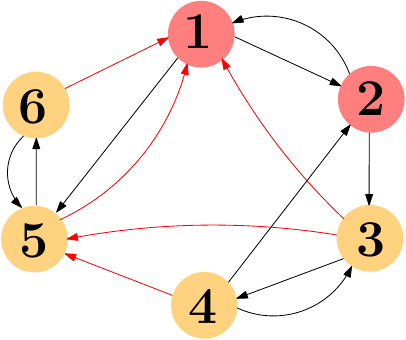}
    \caption{Network $\mathcal{G}$}
    \label{fig:example_netw}
    \end{subfigure}
        \begin{subfigure}{0.24\textwidth}
        \centering
            \includegraphics[width=0.6\linewidth]{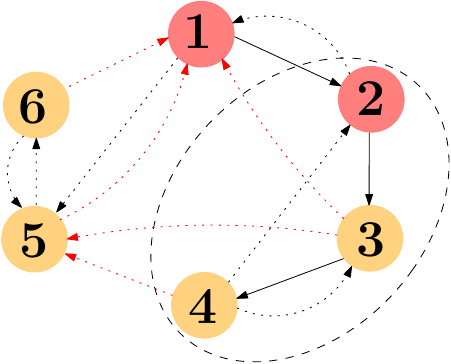}
    \caption{The LTP agent $2$ with $\mathcal{N}_2=\{3,4\}$}
    \label{fig:LTP_2}
    \end{subfigure}
        \begin{subfigure}{0.26\textwidth}
        \centering
            \includegraphics[width=0.55\linewidth]{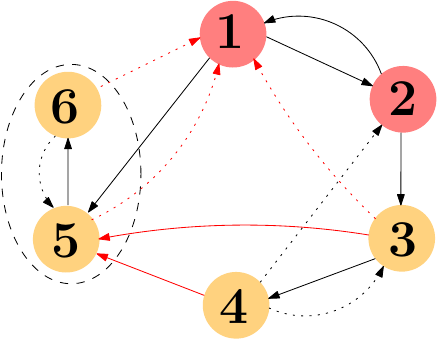}
    \caption{The LTP agent $5$ with $\mathcal{N}_5=\{6\}$}
    \label{fig:LTP_5}
    \end{subfigure}
    \caption{The LTP agents in $\mathcal{G}$.} 
    \label{fig:placeholder}
\end{figure}

\begin{expm}
Consider a signed network shown in Fig. \ref{fig:example_netw} with stubborn agents $1$ and $2$. Throughout this paper, the red nodes denote the stubborn agents and the orange nodes represent the non-stubborn ones. The black arrows denote cooperative interactions, and the red arrows represent the antagonistic ones. Fig. \ref{fig:LTP_2} shows that each path from the stubborn agent $1$ (highlighted by solid lines) to the non-stubborn agents $3$ and $4$ traverses $2$. Since $2$ is a stubborn, it is an LTP agent with $\mathcal{N}_2=\{3,4\}$. Similarly, Fig. \ref{fig:LTP_5} illustrates that each path from stubborn agents $1$ and $2$ to the non-stubborn agent $6$ traverses $5$. Additionally, the paths from the two stubborn agents to $5$ do not traverse any common node. Hence, both conditions (i) and (ii) in Defn. \ref{Definition:LTP_agents} hold for the non-stubborn agent $5$, making it an LTP agent with $\mathcal{N}_5=\{6\}$. 
\end{expm}

An LTP agent has the following properties: 
\begin{lemma}[\cite{shrinate2025opinionclustering}]
Consider a network $\mathcal{G}$ such that $i$ and $j$ are LTP agents. Then, $\mathcal{N}_i \cap \mathcal{N}_j=\emptyset$.     
\end{lemma}

\begin{remark}
\label{rm:partition}
Let a digraph $\mathcal{G}$ have a set of LTP agents denoted by $\mathcal{L}$ such that $\cup_{j\in \mathcal{L}} \{j\} \cup \mathcal{N}_j =\mathcal{V}$. Since the persuaded sets $\mathcal{N}_i$ and $\mathcal{N}_j$ for two distinct LTP agents $i$ and $j$ are disjoint, the LTP agents together with their persuaded sets form a partition of $\mathcal{G}$. 
\end{remark}
\begin{lemma}
\label{lm:neighbours_LTP}
If an agent $p$ is an LTP agent and $q\in \mathcal{N}_p$, then each in-neighbour of $q$ belongs to the set $\mathcal{N}_p\cup \{p\}$    
\end{lemma}
\begin{pf}
Suppose $q$ has an in-neighbour $t\notin \mathcal{N}_p\cup \{p\}$. By definition, each path from every stubborn agent in $\mathcal{G}$ to $q$ traverses $p$. As a result, it follows that the paths from each stubborn agent to $t$ must also traverse $p$. Thus, $t\in \mathcal{N}_p$, which contradicts our assumption that $t\notin \mathcal{N}_p$. 
\end{pf}
\section{Topological Conditions for Multiconsensus under FJ}
\label{Sec:Main_Res}
\vspace{-8pt}
In this section, we begin by relating the final opinions of an LTP agent and those persuaded by it. 
Thereafter, we use this relation to derive conditions under which the LTP agents can result in multiconsensus in the signed network.

Consider a signed network $\mathcal{G}$ with $n$ agents out of which $m$ are stubborn such that Assumption \ref{Assumption:1} holds. For such a network $\mathcal{G}$, the opinions of agents evolving under the SFJ model \eqref{eqn:Signed_FJ_m} always converge and the final opinions $\mathbf{x}^*$ satisfy the following eqn.
\begin{align}
\label{eqn:steady_2}
\mathbf{x}^* = (I_n-\beta)W \mathbf{x}^*+\beta \mathbf{x}(0)    
\end{align}
We can renumber the agents such that the agents $1,...,m$ are stubborn and the rest are non-stubborn. Thereafter, we rewrite eqns. \eqref{eqn:steady_2} as:
\begin{align}
\label{eqn:steady_state}
\underbrace{\begin{bmatrix}
 I_n-(I_n-\beta)W & -\beta[n,m] \\
    \mathbf{0}_{n \times m} & \mathbf{0}_m
\end{bmatrix}}_{R} \
\underbrace{
\begin{bmatrix}
    \mathbf{x}^*\\
    \mathbf{x}_s(0)
\end{bmatrix}}_{\mathbf{y}}=\mathbf{0}
\end{align}
where $\mathbf{x}_s(0)$ denotes the initial opinions of $m$ stubborn agents.
The matrix $R=[r_{i,j}]\in \mathbb{R}^{(n+m)\times (n+m) }$ and vector $\mathbf{y}\in \mathbb{R}^{n+m}$ are defined in eqn. \eqref{eqn:steady_state}. We can say that the first $n$ rows of $R$ associate with the final opinions of the $n$ agents in $\mathcal{G}$, and the last $m$ rows associate with the initial opinions of the stubborn agents. 
\begin{remark}
If an incoming edge to an agent $i$ in $\mathcal{G}$ has a negative weight, then the $i^{th}$ row-sum of $R$ is strictly positive. However, if each incoming edge to $i$ has a positive weight, the $i^{th}$ row-sum of $R$ is equal to $0$.  
\end{remark}

To determine the relation between the final opinions of an LTP agent $p$ and an agent $q \in \mathcal{N}_p$, 
we reduce the linear equations in \eqref{eqn:steady_state} using Schur complement \eqref{eqn:Schur_complement}. 
Next, we show that for suitable choice of $\alpha\subseteq[n+m]$, the matrix $R[\alpha^c]$, required to compute the Schur complement $R/\alpha^c$, is invertible.
\begin{lemma}
\label{lm:Schur_complement_R}
Consider a signed network $\mathcal{G}$ such that Assumption \ref{Assumption:1} holds. Let the opinions of agents in $\mathcal{G}$ evolve under the FJ model \eqref{eqn:Signed_FJ_m} and $R$ be the matrix obtained from eqn. \eqref{eqn:steady_state}. If $p$ is an LTP agent and $q\in \mathcal{N}_p$, then for $\alpha=[n+m]\setminus (\mathcal{N}_p \setminus \{q\})$ and $\alpha^c=\mathcal{N}_p \setminus \{q\}$, 
the submatrix $R[\alpha^c]$ is invertible and is given by
\begin{align}
\label{eqn:R_alpha_inverse}
   R[\alpha^c]^{-1}=\sum_{k=0}^{\infty}W[\alpha^c]^k
\end{align}
\end{lemma}
\begin{pf}
Since each node in $\mathcal{N}_p$ is non-stubborn, 
$R[\alpha^c]=I-W[\alpha^c]$. Let $\tilde{W}_{\alpha^c}=\operatorname{abs}(W[\alpha^c])$. By Lemma \ref{lm:spectral_positive}, we know that $\rho(W[\alpha^c])\leq \tilde{W}_{\alpha^c}.$  Therefore, we show that $\rho(W[\alpha^c])<1$ by establishing that $\rho(\tilde{W}_{\alpha^c})<1$. Consequently, $R[\alpha^c]$ is invertible and eqn. \eqref{eqn:R_alpha_inverse} follows by the Neumann series in \cite{bullo}.

To prove that $\rho(\tilde{W}_{\alpha^c})<1$, we use the following properties of $W$ and nodes in $\alpha^c$.

\begin{enumerate}
    \item[(P1)] By definition of $W$, its enteries satisfy 
    $\sum_{j=1}^n |w_{ij}|=1$. 
    \item[(P2)] Under the Assumption \ref{Assumption:1}, and Defn. \ref{Definition:LTP_agents}, each non-stubborn agent in $\alpha^c$ has a path from at least one stubborn agent, and each such path traverses $p$. 
\end{enumerate}
Due to (P1), the row sum of each row in $\tilde{W}_{\alpha^c}$ is at most one. Further, by (P2), there exists an agent in $\alpha^c$ whose in-neighbour is either $p$ or $q$ (which are not in $\alpha^c$). Together, (P1) and (P2), imply that $\tilde{W}_{\alpha^c}$ is row-substochastic. 

Additionally, by Assumption \ref{Assumption:1}, each non-stubborn agent $i \in \alpha^c$ with row-sum in $\tilde{W}_{\alpha^c}$ equal to $1$ has a path in $\mathcal{G}$ from an agent $j\in \alpha^c$ with a corresponding row-sum strictly less than $1$. Clearly, this property also holds for the associated graph $G(\tilde{W}_{\alpha^c})$. Hence, $\rho(W[\alpha^c])<1$ by Theorem 4.11 in \cite{bullo}. \qed
\end{pf}

By Lemma \ref{lm:Schur_complement_R}, it follows that for an LTP agent $p$ and $q\in \mathcal{N}_p$, we can evaluate the Schur complement of $R$ for $\alpha^c=\mathcal{N}_p\setminus \{q\}$. It is given as:
\begin{align}
\label{eqn:Schur_Complement_R_a}
    R/\alpha^c=R[\alpha]+R[\alpha,\alpha^c](\sum_{k=0}^{\infty}W[\alpha^c]^k)R[\alpha^c,\alpha]
\end{align}
It follows from \eqref{eqn:reduced_LE} that we can use the Schur complement $R/\alpha^c$ to reduce the linear eqn. \eqref{eqn:steady_state} as follows:
\begin{align}
\label{eqn:reduced_R_alpha}
    R/\alpha^c \cdot \mathbf{x}^*[\alpha]= \mathbf{0}_{|\alpha|}
\end{align}
Since nodes $p,q \in \alpha$, eqn. \eqref{eqn:reduced_R_alpha} establishes the relation 
between final opinions of $p$ and $q \in \mathcal{N}_p$. We use this relation in the following result to determine the conditions under the LTP agent $p$ forms an opinion cluster.

\begin{figure}[ht]
    \centering
    \begin{subfigure}[c]{0.42\linewidth}
        \centering
        \includegraphics[width=0.9\linewidth]{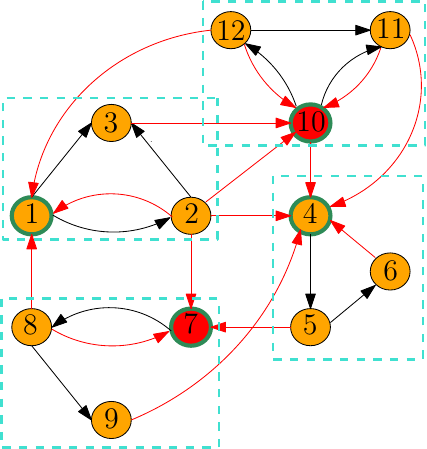}
        \caption{\footnotesize Network $\mathcal{G}$}
        \label{fig:opinion_clusters_network}
    \end{subfigure}
    \hfill
    \begin{subfigure}[c]{0.545\linewidth}
        \centering
        \includegraphics[width=0.9\linewidth]{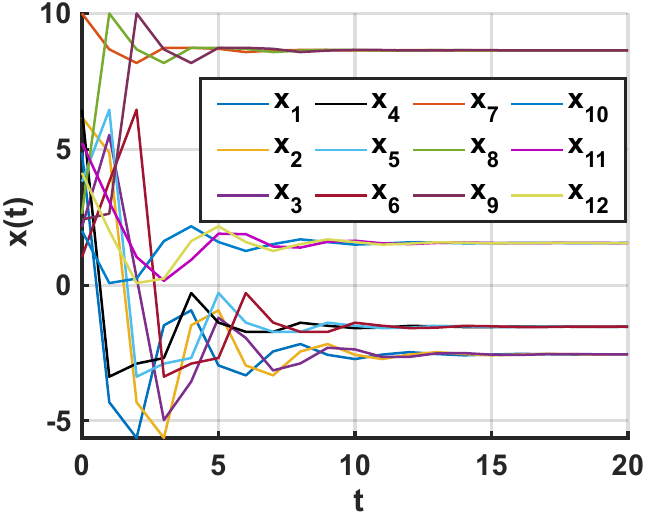}
        \caption{\footnotesize Formation of opinion clusters.}
        \label{fig:opinion_clusters_plot}
    \end{subfigure}
    \caption{\footnotesize The network with LTP agents results in the formation of opinion clusters under the SFJ model.}
    \label{fig:opinion_clusters}
\end{figure}

\begin{thm}
\label{thm:main_result}
Consider a signed network $\mathcal{G}$ with $m>1$ stubborn agents such that Assumption \ref{Assumption:1} holds. The opinions of agents are evolving under the SFJ model \eqref{eqn:Signed_FJ_m}. If $p$ is an LTP agent and all incoming edges at each node in $\mathcal{N}_p$ are cooperative, then $p$ and nodes in $\mathcal{N}_p$ form an opinion cluster.  
\end{thm}
\begin{pf}
Let $q \in \mathcal{N}_p$. Our objective is to show that $x_p^*=x_q^*$. Consider the case where $|\mathcal{N}_p|\geq 2$, the remaining case follows similarly. Under this case, let $\alpha^c=\mathcal{N}_p \setminus \{q\}$ and $\alpha=[n+m]\setminus \alpha^c$. 

Suppose the agents are re-numbered such that $1,...,|\alpha|$ are the nodes in set $\alpha$ with nodes $1$ and $2$ being $p$ and $q$, which are followed by the remaining nodes in set $\alpha^c$. From Lemma \ref{lm:neighbours_LTP}, we know that none of the nodes in set $\alpha$ except $p$ can be an in-neighbour of $q$. Therefore, $R[\alpha]$ has the following form:
\begin{align*}
    R[\alpha]=\begin{bmatrix}
        r_{1,1} & r_{1,2} & r_{1,3} & ... & r_{1,|\alpha|} \\
        r_{2,1} & r_{2,2} & 0 & ...& 0 \\
        r_{3,1} & r_{3,2} & r_{3,3} & \cdots & r_{3,|\alpha|}\\
        \vdots & \vdots & \cdots& \cdots& \vdots \\
        r_{|\alpha|-m,1} & r_{|\alpha|-m,2} & r_{|\alpha|-m,3}& \cdots & r_{|\alpha|-m,|\alpha|} \\
        \mathbf{0}_{m} &  \mathbf{0}_{m} &  \mathbf{0}_{m}&  \cdots & \mathbf{0}_{m}
    \end{bmatrix}
\end{align*}

Next, we evaluate the Schur complement $R/\alpha^c$ given by \eqref{eqn:Schur_Complement_R_a}.
The matrix $R/\alpha^c=[r_{i,j}^1]$ has $|\alpha|$ rows and columns corresponding to the nodes in set $\alpha$. Let $F:\alpha \to |\alpha|$ denote a mapping of nodes in set $\alpha$ to the rows of matrix $R/\alpha^c$ such that $F(p)=1$ and $F(q)=2$. From eqn. \eqref{eqn:Schur_Complement_R_a}, it follows that an entry $r_{F(i),F(j)}^{1}$ is non-zero only when either $r_{i,j}$ is non-zero or there is a path $j$ to $i$ in $\mathcal{G}$ that passes only through nodes in $\alpha^c$ (except $i$ and $j$). 

By the definition of LTP agents and Lemma \ref{lm:neighbours_LTP}, a path from any agent $c \notin \mathcal{N}_p$ to $q$ always passes through $p$. Consequently, $e_2^T R/\alpha^c=\begin{bmatrix}
  r_{2,1}^1 & r_{2,2}^1 & \mathbf{0}_{|\alpha|-2}  
\end{bmatrix}$, where $e_2$ is the standard basis vector with second entry equal to $1$.

Next, we know that if an agent $i$ has only cooperative interactions the corresponding row-sum in $R$ is zero. Given that each node in $\mathcal{N}_p$ only has cooperative interactions, we claim that the row in $R/\alpha^c$ corresponding to node $q$ also has row-sum equal to $0$. 

To prove our claim, we use the following relation satisfied by $R$ as $\alpha^c \subset \mathcal{N}_p$
\begin{align}
    \begin{bmatrix}
        R[\alpha] & R[\alpha,\alpha^c] \\
         R[\alpha^c,\alpha] &  R[\alpha^c]
    \end{bmatrix}\begin{bmatrix}
        \mathbf{1}_{|\alpha|
        } \\ \mathbf{1}_{|\alpha^c|}
    \end{bmatrix}=\begin{bmatrix}
        \mathbf{u} \\ \mathbf{0}
    \end{bmatrix}
\end{align}
where $\mathbf{u}=[u_1,...,u_{|\alpha|}]$ has non-negative enteries with $u_2=0$ because second row of $R$ corresponds to node $q$ and all the incoming edges to $q$ are cooperative. By eqn. \eqref{eqn:reduced_LE}, we get that
\begin{align*}
    R/\alpha^c\mathbf{1}_{|\alpha|}=\mathbf{u}
\end{align*}
Since $u_2$ is zero, the second row of $R/\alpha^c$ corresponding to $q$ has row-sum equal to $0$. Therefore, $r_{2,1}^1=-r_{2,2}^1$. Consequently, from eqn. \eqref{eqn:reduced_R_alpha}, it follows that $x_p^*=x_q^*$. \qed

\begin{corollary}
\label{Cor:Clustering_bal}
Consider a signed network $\mathcal{G}$ with $m>1$ stubborn agents such that Assumption \ref{Assumption:1} holds. The opinions of agents evolving under the SFJ model \eqref{eqn:Signed_FJ_m} reach multiconsensus with $z \geq m$ clusters if  the following conditions hold:
\begin{enumerate}
    \item[(C1)] $\mathcal{G}$ has $z$ LTP agents denoted by set $\mathcal{L}$ such that $\cup_{j\in \mathcal{L}} {j} \cup \mathcal{N}_j=\mathcal{V}$.
    \item[(C2)] Each incoming edge at agent $i \in \mathcal{N}_j$ is cooperative  $\forall i\in \mathcal{N}_j$  and $\forall j \in \mathcal{L}$.
\end{enumerate}
 
\end{corollary}
\begin{pf}
Remark \ref{rm:partition} shows that the LTP agents and their persuaded sets form a partition of the network $\mathcal{G}$ with each subgroup composed of an LTP agent and the agents it persuades. Thereafter, Theorem \ref{thm:main_result} establishes that under the given conditions an LTP agent and the agents it persuades form an opinion cluster. Thus, multiconsensus occurs under the SFJ model.  \qed
\end{pf}
Corollary \ref{Cor:Clustering_bal} presents the graph topology-based conditions that ensure multiconsensus in signed digraphs. Under the achieved multiconsensus, each opinion cluster is formed by an LTP agent $p \in \mathcal{L}$ and the agents in $\mathcal{N}_p$ provided that the condition (C2) holds. Together with Lemma \ref{lm:neighbours_LTP}, condition (C2) ensures that the incoming edges to each agent $\mathcal{N}_p$ from the agents $p$ and $\mathcal{N}_p$ are cooperative. This aligns with existing literature on multiconsensus  \cite{xia2011clustering,QIN20132898,luo2021cluster,maria}, where intracluster interactions are assumed to be cooperative. However, unlike standard assumptions, an LTP agent $p$ can have incoming edges with negative weights  from the nodes in $\mathcal{N}_p$ under
(C2). 


A key advantage of the proposed conditions is that they are independent of the edge weights and the stubbornness values of the agents. Thus, the proposed conditions can lead to multiconsensus in the graphs that do not satisfy in-degree balance (e.g. graphs with clustering balance) can achieve multiconsensus. Further, Corollary \ref{Cor:Clustering_bal} generalises the conditions for multiconsensus presented in \cite{maria} for undirected graphs to weakly connected digraphs. 

\begin{remark}
The conditions for multiconsensus in Corollary \ref{Cor:Clustering_bal} are sufficient conditions because for specific values of the edge weights there may exist networks that do not have LTP agents but satisfy eqn. \eqref{eqn:V_enteries}. However, designing edge weights for such networks is complex. On the contrary, the proposed topology based conditions are simple to implement and robust to perturbations in edge weights 

\end{remark}

The following example demonstrates the results presented in Corollary \ref{Cor:Clustering_bal}.

\end{pf}
\begin{expm}
 Consider the network $\mathcal{G}$ shown in Fig. \ref{fig:opinion_clusters_network} with stubborn agents $7$ and $10$. The agents $1,4,7$ and $10$ in $\mathcal{G}$ are the LTP agents (highlighted with green boundary in Fig. \ref{fig:opinion_clusters_network}) such that $\mathcal{N}_1=\{2,3\}$, $\mathcal{N}_4=\{5,6\}$, $\mathcal{N}_7=\{8,9\}$ and $\mathcal{N}_{10}=\{11,12\}$. Together, the LTP agents and the agents persuaded by them form a partition of the given network such that each cluster is highlighted by the blue dashed boxes. Note that each agent that is persuaded by an LTP agent only has cooperative interactions. Therefore, for initial conditions chosen over a uniform distribution over $[0,10]$ and stubbornness values in $[0,1]$, the opinions of the agents evolving by the SFJ model \eqref{eqn:Signed_FJ_m} converge to $4$ opinion clusters shown in Fig. \ref{fig:opinion_clusters_plot}.
   
\end{expm}

\vspace{-7pt}
\section{Conclusion}
\label{Sec:Conclusion}
This work analyses the topological properties of signed networks to achieve multiconsensus. Leveraging the notion of LTP agents, we establish the sufficient conditions under which multiconsensus emerges under the opposing-rule SFJ model. Our results demonstrate that the SFJ model can serve as an effective distributed control strategy for multiconsensus in signed digraphs. Moreover, the proposed conditions  
broaden the class of signed networks for which multiconsensus can be achieved as compared to \cite{maria}.
In future, we plan to develop an algorithm that achieves any desired multiconsensus. 




\vspace{-7pt}

\bibliography{ifacconf}             
                                                   







\appendix
\end{document}